\documentclass[aps,prl,twocolumn,groupedaddress]{revtex4-1}

\usepackage{epsfig}
\usepackage{amsmath}
\usepackage{graphicx}
\usepackage{dcolumn}
\usepackage{latexsym}
\usepackage{color}
\usepackage{epstopdf}
\usepackage{subfigure}
\usepackage{comment}
\usepackage{nicefrac}
\usepackage{braket}
\usepackage[ulem=normalem]{changes}
\usepackage{float}
\usepackage{xcolor}
\usepackage{soul}

\begin{document}

\title{Quantum criticality at the superconductor insulator transition probed by the Nernst effect}

\author{A. Roy}
\author{E. Shimshoni}
\author{A. Frydman}

\affiliation{Department of Physics, Bar Ilan University, Ramat Gan 52900, Israel}

\date{\today}

\begin{abstract}
The superconductor-insulator transition (SIT) is an excellent example for a quantum phase transition at zero temperature, dominated by quantum fluctuations. These are expected to be very prominent close to the quantum critical point. So far most of the experimental study of the SIT has concentrated on transport properties and tunneling experiments which provide indirect information on criticality close to the transition. Here we present an experiment uniquely designed to study the evolution of quantum fluctuations through the quantum critical point. We utilize the Nernst effect, which has been shown to be effective in probing superconducting fluctuation. We measure the Nernst coefficient in amorphous indium oxide films tuned through the SIT and find a large signal on both the superconducting and the insulating sides which peaks close to the critical point. The transverse Peltier coefficient, $\alpha_{xy}$ which is the thermodynamic quantity extracted from these measurements, follows quantum critical scaling with critical exponents $\nu\sim0.7$ and $z\sim1$ which is consistent with a clean XY model in 2+1 dimensions. 
\end{abstract}


\maketitle

Quantum fluctuations are crucial for understanding fundamental physics from atomic scale to the scale of the universe. Most prominently, they are the driving force behind a quantum phase transition (QPT) between two competing phases of matter at zero temperature \cite{QPTbook}.  An experimentally versatile example for a QPT is the supurconductor-insulator-transition (SIT) in thin superconducting films, which is driven by quantum fluctuations and controlled by a non thermal tuning parameter $g$. For $g < g_c$ the film is a superconductor and for  $g > g_c$  the system becomes insulating. Experimentally, different $g$'s have been  used to drive the transition including inverse thickness \cite{Strongin1970,Dynes1986,Haviland1989,Valles1992,Frydman2002,Hadacek2004,Stewart2007,Sacepe2008,Hollen2011,Postolova2017,Baturina2011,Poran2017}, magnetic field \cite{Hadacek2004,Stewart2007,Paalanen1992,Yazdani1995,Gantmakher1998,Sambandamurthy2004,Sambandamurthy2005,Steiner2005,Baturina2005,Baturina2007,Crane2007,Vinokur2008,Ganguly2017}, disorder \cite{Crane2007,Shahar1992,Sacepe2011,Poran2011}, chemical composition \cite{Mondal2011} and gate voltage \cite{Parendo2005,Caviglia2008,Bollinger2011}. Though the quantum critical point ($g=g_c$) occurs at zero temperature,  it also profoundly affects the behavior of the system at finite temperature. In the quantum critical regime the system is neither superconducting nor insulating, and is dominated by quantum fluctuations. These fluctuations of the superconducting order parameter, $\psi=\psi_0  e^{i\theta}$, can be both amplitude ($\psi_0$) and phase ($\theta$) fluctuations which are interrelated via the uncertainty principle.

\begin{figure}[h]
\vspace{0cm}
\centering
\includegraphics[width=0.43\textwidth]{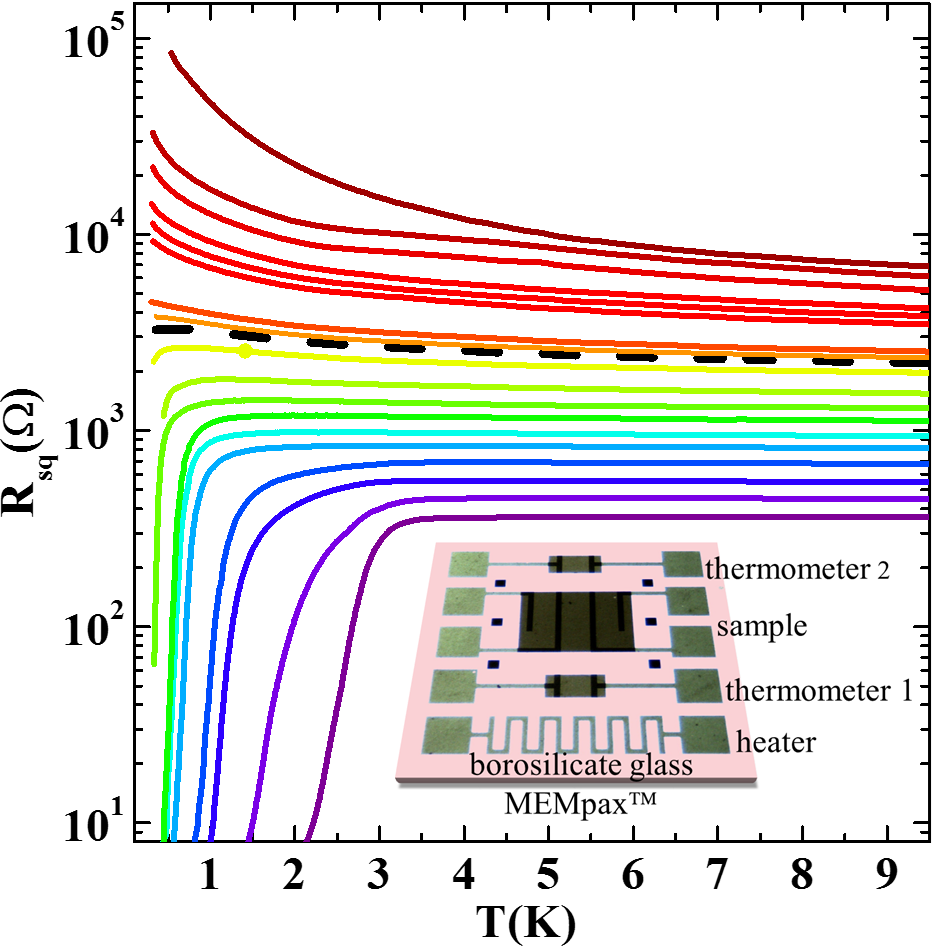}
\caption{(color online). $R_{\Box}$ vs. T for different annealing stages. The quantum phase transition is manifested as the gradual change of ground state from insulator to superconductor as $R_{\Box}$ is lowered. The dashed line is the curve for the film characterized by  $g\simeq g _c$, $g_c$ being the value extracted from the scaling analysis below. The curve separates the insulating and the superconducting stages. Inset: Optical image of the chip containing an Au meander as a heater, two strongly insulating films utilized as thermometers and the InO$_x$ sample . About 1/4 of the chip near thermometer 2 is anchored thermally to the cold head. }
\label{setup}
\end{figure}

While much progress has been made in the field over the years both theoretically and experimentally, there are still important open questions concerning phenomena close to the SIT. In particular, to this date it remains controversial which universality class best describes the observed transitions, and to what extent it varies between different specific realizations. From a theoretical point of view, a prototypical model which captures quantum fluctuations in $\psi$ can be cast in terms of repulsively interacting Bosons such as the Bose-Hubbard model, or equivalently an array of Josephson-coupled superconducting islands where a charging energy $E_C$ competes with the Josephson energy $E_J$ \cite{MatthewFisher1990,ES_Fisher_Fisher89_Fisher1989,ES_SITrev_Sondhi1997}.
This introduces a natural tuning parameter, e.g. $g=E_C/E_J$. However, a generic model of relevance to the physical system involves additional parameters which may profoundly affect the SIT: a chemical potential (which tunes the occupation of Bosons per site), a magnetic field, and disorder, introduced as randomness in all the above parameters.
 This suggests a variety of quantum critical points with distinct critical behavior, manifested by different possible values of the critical exponents characterizing, e.g., the divergence of the correlation length ($\xi$) and time ($\xi_\tau$) with the deviation from the quantum critical point $\Delta g=g-g_c$: \cite{QPTbook}
\begin{equation}
\xi\sim|\Delta g|^{-\nu}\; ,\quad \xi_\tau\sim\xi^z\; .
\label{nu_z_def}
\end{equation}

In the clean limit, the insulating phase is interaction-dominated (a Mott insulator). The dynamical critical exponent $z$ depends on the commensurability of Boson occupations and is either $z=1$, if particle-hole symmetry is obeyed, or $z=2$ at generic filling. In the former case, the SIT can be mapped to the classical 3D XY-model, yielding $\nu\approx 2/3$ \cite{ES_3DXY_Li1989,ES_3DXY_Campostrini2001}.
Disorder introduces an intermediate, gapless insulating phase dubbed \textquotedblleft Bose glass" \cite{ES_Fisher_Fisher89_Fisher1989}
which undergoes a direct transition to a superfluid. The critical exponents were argued to be $z=d$ (i.e. $z=2$ in 2D), and $\nu\geq 1$, whereas long-range Coulomb interactions imply $z=1$ \cite{ES_FGG_Fisher1990}.
Extensive numerical works over the past two decades \cite{ES_ChaGirvin_Sørensen1992,ES_ChaGirvin_Cha1994,ES_Prokoviev_Prokofev2004,ES_IPR_Iyer2012,ES_Vojta_Vojta2016}, addressing arbitrarily large disorder strength and the role of magnetic field, have yielded estimates of $1<z<2$ (e.g. $z=1.52$ in \cite{ES_Vojta_Vojta2016}), and various values of $\nu$ consistent with the bound $\nu\geq 1$.

On the experimental front, so far attempts to provide the critical exponents were based on dc transport via scaling analyses of resistivity data\cite{Hebard1990,Yazdani1995,Theunissen1997,Markovic1998,Markovic1999,Gantmakher2000,Bielejec2002,Aubin2006,Caviglia2008,Bollinger2011,Shi2014,Park2017,Marrache-Kikuchi2008}. Typically, these experimental results are consistent with $z=1$, but the reported values of $\nu$ range from $0.4$ to $2.3$ and do not obviously agree with the theoretical predictions. Indeed, resistivity is possibly not an ideal probe of critical fluctuations in the order parameter field, since it is sensitive to details such as the specific scattering mechanism, inhomogenieties etc. Moreover, it is not a thermodynamic quantity and is inherently non-equilibrium. Quantum fluctuations close to the SIT have been observed in thermodynamic measurements, e.g. of specific heat  \cite{Poran2017} and susceptibility \cite{Mondal2011}. However, these have not provided quantitative information on the critical behavior.

A promising candidate for fluctuation studies is the Nernst effect, i.e. the appearance of a transverse electric field in the presence of a longitudinal thermal gradient and a perpendicular magnetic field \cite{Behnia2009,Behnia2016,Ussishkin2002,Podolsky2007,Hartnoll2007,Michaeli2009,Wachtel2014}. In recent years a substantial Nernst signal $N=E_y/(-\nabla_xT)$ was measured around and above the critical temperature $T_c$ in the underdoped regime of high-Tc superconductors \cite{Wang2006} and in 2D disordered (NbSi and InO$_x$) films \cite{Pourret2007,Spathis2008}. In the latter it was shown that the unexpectedly large Nernst effect is due to the motion of vortices above the Berezinskii-Kosterlitz-Thouless  temperature T$_{BKT}$. However, up to date there has been no experimental study of the Nernst effect throughout the entire SIT.

Recent theoretical studies on quasi 1D Josephson junction arrays predict a pronounced peak of N close to the SIT due to quantum phase fluctuations \cite{Atzmon2013,Schattner2016}. This peak grows as the temperature is lowered towards T = 0. A qualitatively similar behavior was predicted to hold for a 2D system as well.

In this Letter we describe a comprehensive measurement of the Nernst effect on an amorphous indium oxide (InO$_x$) film driven continuously through a disorder-induced SIT. This enables us to extract a thermodynamic quantity, the off-diagonal Peltier coefficient $\alpha_{xy}$, in order to quantitatively explore the quantum criticality.  Our main findings are as follows:

\begin{itemize}
\item{Sizable Nernst signal is measured on both the superconducting and the insulating sides of the disorder-driven SIT}
\item{The Nernst effect amplitude peaks close to the SIT in accordance with recent theoretical predictions. The maximum occurs at $g \simeq 0.35g_c$ }
\item{$\alpha_{xy}$ exhibits data collapse over many orders of magnitude providing a direct determination of the universality class of quantum fluctuations close to the SIT. The scaling analysis is consistent with a clean (2+1)D-XY model yielding critical exponents $\nu\sim$0.7 and z$\sim$1.}
\end{itemize}

An InO$_x$ film of thickness 30nm was e-beam evaporated on MEMpax$\texttrademark$ borosilicate glass substrate of thickness 0.4mm. This substrate was chosen due to its very low thermal conductivity at low temperatures. An Au meander utilized as a heater and two on-chip thermometers (strongly insulating InO$_x$ films) were also evaporated in order to allow a Nernst-effect setup as shown in Fig.\ref{setup}(Inset). Thermal contact with a 330mK $^3$He cryostat was provided at the edge of the substrate farthest from the heater, which determined the direction of the heat current. DC measurements of the transverse thermoelectric voltage and the resistance were carried out with a Keithley 182 digital voltmeter.

\begin{figure}
\vspace{0cm}
\centering
\includegraphics[width=0.44\textwidth]{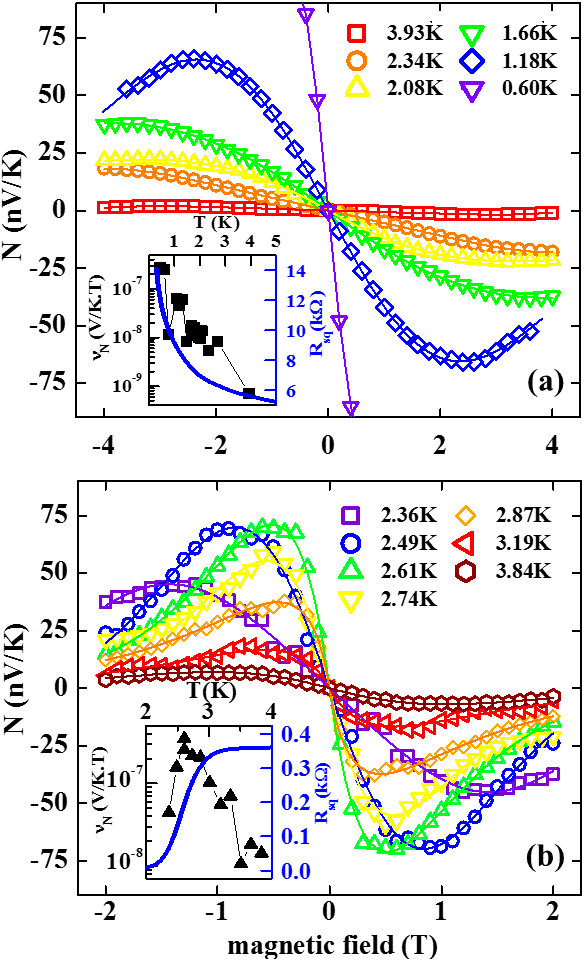}
\caption{(color online). Nernst signal, N, versus magnetic field at various temperatures, deep in the insulating ($g/g_c=2.16$; panel a)  and the superconducting ($g/g_c=0.15$; panel b) sides of the SIT. Also shown are the fitted curves with the ad-hoc analytic function for extraction of the Nernst coefficient $\nu _N$. Insets: extracted Nernst coefficient $\nu$ vs T (symbols) and the respective $R_{\Box}$(T) curves (blue lines).}
\label{nernst}
\end{figure}

The transformation from an insulating ground state to a superconducting one was carried out by increasing the electrical conductance of the sample in stages via low temperature thermal annealing \cite{ZvieOvadyahu1986}. An initial highly resistive sample (R$_{\Box 5K} \simeq$10$k\Omega$) was created using a high O$_2$ partial pressure (8$\times$10$^{-5}$ Torr) during evaporation. It was then taken through several cycles of annealing and measurement, decreasing the room-temperature resistance by $\approx$5-10$\%$ in each cycle. Resistance versus temperature for the different annealing stages is presented in Fig.\ref{setup}. The tuning parameter, $g$, was chosen to be the sheet resistance, R$_{\Box}$ at $T=5K$ in units of $h/4e ^2$. From the data analysis detailed below yields we find $R^{c}_{\Box 5K} = 2410 \Omega$. This value yields the dashed line in Fig.1 which separates insulating and superconducting curves. The obtained T$_c$'s for the succeeding stages showed a monotonic increase with decreasing $g=R_{\Box 5K} \cdot 4e ^2/h$.

For each annealing stage the Nernst signal was measured as a function of magnetic field at different temperatures in the range of 0.4K to 4K. In every case, including those in the insulating phase, the field dependence of the signal showed features similar to that of a typical superconductor as reported elsewhere \cite{Behnia2016}: an asymmetric peak, whose position shifted with temperature. The Nernst coefficient $\nu _N$ = N/B in the  limit B$\rightarrow$0 was extracted by fitting the data with an ad-hoc fitting function N(B)= $\nu _N$Be$^{-\mu \mid B \mid ^c}$. Fig. \ref{nernst} depicts such measurements for two annealing stages, one deep in the insulating phase and the other deep in the superconducting phase. It is seen that the overall Nernst features are similar for the two phases, though the temperature dependence is slightly different. For the superconducting stage $\nu _N$ exhibits a peak near the mean field $T_c$ while for for the insulating stage (that obviously does not have a finite $T_c$) $\nu _N$ shows monotonic decrease over several orders of magnitude.

\begin{figure}
\vspace{0cm}
\centering
\includegraphics[width=0.47\textwidth]{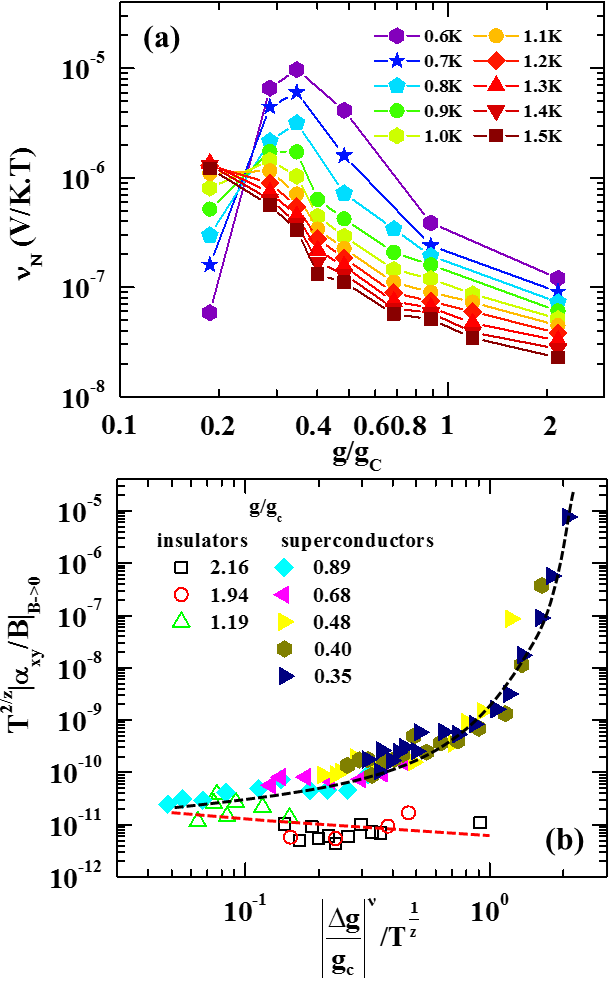}
\caption{(color online). (a) $\nu _N$ versus the normalized quantum tuning parameter, $g/g_c$ showing a peak at $g < g_c$, in agreement with \cite{Atzmon2013}. (b):Scaling plot of the off-diagonal Peltier coefficient $\alpha _{xy}$. Best data collapse was found for critical exponents $\nu = 0.70$ and  $z = 0.99$ and critical resistance $R^{c}_{\Box 5K}=2410\Omega$. $g=R _{\Box 5K} \cdot 4e ^2 /h$. The dashed lines are guides to the eye.}
\label{alpha}
\end{figure}

Fig.\ref{alpha}a shows $\nu _N$ as a function of $g/g_c$ through the SIT for different temperatures. At low temperature $\nu _N$ peaks  at $g \simeq 0.35g_c$. The peak amplitude decreases as the temperature is increased. This is consistent with theoretical predictions \cite{Atzmon2013} and has been attributed to the fact that in systems with an effective particle-hole symmetry, the Nernst signal can be generally expressed as a product of the resistivity and the transverse Peltier coefficient $N=\rho_{xx} \cdot \alpha_{xy}$. The non-monotonous behavior of $\nu _N$  arises from the  competition between $\rho_{xx}$, which increases with $g$, and $\alpha_{xy}$, which signifies the strength of the diamagnetic response and hence decreases with $g$.

We note that $\rho_{xx}$ is a non-equilibrium property signifying the rate of phase slips, and is therefore relatively sensitive to microscopic details. In contrast, $\alpha_{xy}$ is approximately proportional to the diamagnetic moment\cite{Podolsky2007,Hartnoll2007,Wachtel2014,Atzmon2013,Schattner2016}, i.e. it is a thermodynamic quantity and is expected to be dominated by universal properties. In order to isolate the thermodynamic contribution of the Nernst effect we extract $\alpha_{xy} =N/R_{\Box}$ for each $g$ and $T$ of figure \ref{alpha}a calculated using the temperature dependent $\rho_{xx}$. Since the critical behavior is expected to hold only in the immediate neighborhood of $g_c$ we focus on samples in the regime $-0.65 < \frac{\Delta g}{g_c} < 1.2$ since in this regime the analysis described below yielded consistent results (see supplementary material). These are plotted in figure \ref{alpha}b using a scaling ansatz, which assumes proximity to a quantum critical point characterized by critical exponents of Eq. (\ref{nu_z_def}). At finite $T$, universal properties are then expected to depend on $g,T$ via the ratio $\xi_\tau/L_\tau$ where $L_\tau\sim 1/T$ is the effective size in the time axis. 

To derive the implied scaling form of $\alpha_{xy}$, we recall the definition
\begin{equation}
\alpha_{xy}=\frac{J_e}{\nabla T}
\label{alpha_xy_def}
\end{equation}
where the electric current $J_e$ has the physical dimension
\begin{equation}
[J_e]\sim {\rm Time}^{-1}{\rm Length}^{-(d-1)}\; .
\end{equation}
Its dependence on $T$, $B$, $\Delta g$ and $\nabla T$ therefore assumes the general form \cite{BGS}
\begin{equation}
J_e(T,B,\Delta g,\nabla T)\sim T^{1+(d-1)/z}F_e\left(\frac{B}{T^{2/z}},\frac{|\nabla T|}{T^{1+1/z}},\frac{|\Delta g|^\nu}{T^{1/z}}\right)
\end{equation}
where $F_e$ is a universal scaling function. For small $B$ and $\nabla T$, $\alpha_{xy}$ and hence $F_e$ is linearly dependent on the first two arguments:
\begin{equation}
F_e\sim \frac{B}{T^{2/z}}\frac{|\nabla T|}{T^{1+1/z}}f_e\left(\frac{|\Delta g|^\nu}{T^{1/z}}\right)
\end{equation}
with $f_e(x)$ a single-parameter scaling function. Inserting in Eq.(\ref{alpha_xy_def}), we thus obtain
\begin{equation}
\alpha_{xy}\sim BT^{(d-4)/z}f_e\left(\frac{|\Delta g|^\nu}{T^{1/z}}\right)=\frac{B}{T^{2/z}}f_e\left(\frac{|\Delta g|^{\nu z}}{T}\right)
\label{alpha_xy_scaling}
\end{equation}
where in the last step we used $d=2$.

For determining the critical exponents we fit the experimental values of   $\alpha_{xy}$ for different T and g to Eq. (\ref{alpha_xy_scaling}). The search for the best data collapse was carried out by minimizing the sum of residuals from two 'best-fitting' polynomial curves, one above and one below $g_c$, using $z$, $\nu$ and $R ^{c}_{\Box 5K}$ as fitting parameters. The procedure \cite{NoteX} led to : $z = 0.99 \pm0.01$; $\nu$ = 0.70$\pm$0.09 and $R^{c}_{\Box 5K}$ = 2410$\pm$69$\Omega$. Figure \ref{alpha}b, shows that this fit yields good data collapse over many orders of magnitude.  It is also seen that the scaling form holds on both sides of the QPT, with different forms of the scaling function $f_e$. This result is consistent with a clean (2+1)D XY model where particle-hole symmetry is effectively obeyed. It provides confirmation that the the SIT is  a quantum phase transition driven by interaction-dominated quantum fluctuations of the superconducting order parameter in 2D.

The XY model is in agreement with the so called \textquotedblleft bosonic model" for the SIT \cite{MatthewFisher1990} in which the system can be modelled by an array of sites, each one  characterized by a local superconducting order parameter amplitude and phase and the probability to obtain phase coherence, and hence global superconductivity, depends on the ratio $E_C / E_J$. InO films, despite being morphologically uniform, have been shown to include \textquotedblleft emergent granularity" in the form of superconducting puddles embedded in an insulating matrix\cite{Kowal1994,Kowal2008,Shimshoni1998,Dubi2007,Imry2008,Trivedi1996,Ghosal2001,Bouadim2011}. Hence, local superconductivity can be present in the insulating phase as well.  The bosonic model separates between the mean-field critical temperature $T_c^{mf}$ which sets the Cooper-pair breaking scale, and  $T_{BKT}$ which is related to the proliferation of free vortices whose motion is measured by transport. The finite Nernst effect we observe on the insulator indicates the presence of vortex motion in this phase as well, thus providing further confirmation for the relevance of the XY model to our systems. In this context we note a few earlier observations on InO$_x$ which revealed the presence of superconductivity in the insulator. The role of vortex-like superconducting fluctuations in the insulating phase were demonstrated by measurements of the \textquotedblleft vortex ratchet effect"\cite{Poran2011} and of Little-Parks oscillations \cite{Kopnov2012} on both sides of the SIT.  In addition, tunneling density of states experiments detected the presence of a superconducting energy gap, and thereby Cooper pairing, not only above T$_c$ \cite{Sacepe2011}, but also on the insulating side of the disorder-driven SIT \cite{Sherman2012}. These fluctuations of the phase and amplitude of the order parameter were picked up by our Nernst measurements. Our results indicate that the true critical behavior (which takes over in the limit of long length-scales) is not sensitive to disorder - but rather dominated by a coarse-grained effective model of coupled SC puddles.

We are grateful to I. Volotsenko for technical help and to K. Behnia for useful discussions. This research was supported by the Israel science foundation, grants No. 783/17 (A.F.) and 231/14 (E.S.).

\bibliography{references/nernst-af,references/nernst-es,references/nernst-ar}

\end{document}